# The Geometry of Putting On a Planar Surface

Robert D. Grober

Department of Applied Physics, Yale University, New Haven, CT  06520.

Abstract

This paper explores the geometry of putting in the limiting case of a planar putting surface.  Putts equidistant from the hole originating on an arc spanning ±30 degrees are shown to share a common target point.  Moving around the circle of all equidistant putts, the ensemble of target points map out a small, diamond-shaped structure centered on the fall line directly above the hole.  The position and size of this target diamond for any length putt on a putting surface of any grade and speed is reasonably approximated by a single universal curve.

This understanding suggests a practical methodology for reading putts.  Instead of lining up only the putt-at-hand, the golfer should line up all putts known to share a common target point.  This methodology will increase the probability of choosing the correct target line.

Introduction

This paper explores the geometry of putting in the limiting case of a planar putting surface, which is often a reasonable approximation within 10-15 feet of the hole. The pioneering work of H.A. Templeton [1, 2] demonstrated that all putts equidistant from the hole share a common target point. This target point is directly uphill from the hole. As the length of the putt, grade and/or speed of the green increases, the target point moves further up the fall line. Unfortunately, Templeton's target point blurs into a progressively larger region as the target point moves up the fall line. This paper expands on Templetons' idea by considering the target point of subsets of equidistant putts originating on an arc spanning ±30 degrees. Moving around the circle of equidistant putts, the target points of the subsets map out a small, diamond-shaped structure centered on the fall line directly above the hole. The position and size of this target diamond for any length putt on a putting surface of any grade and speed is described quantitatively by a universal curve.

This understanding suggests a practical methodology for reading putts. Instead of lining up only the putt-at-hand, the golfer should line up several putts in the family of putts equidistant from the hole and on an arc that spans a range approximately ± 30 degrees relative to the putt-at-hand. By considering a family of putts all known to share the same target point, the golfer increases the probability of correctly identifying the target point, and thus the correct target line.

Putt Trajectories

This paper explores the geometry of putting in the limiting case of a planar putting surface, which is often a reasonable approximation within 10-15 feet of the hole. The planar putting surface is characterized by its grade and speed. Grade is the ratio of the rise over the run and is quoted as a percentage. The grade is also the tangent of the angle $\theta$ between the normal to the green and the orientation of gravity. Because the grades of putting greens are relative shallow, $\tan\theta \approx \theta$ is always a reasonable approximation. Speed is characterized in terms of the Stimp speed [3] of the green. In practice, the Stimp speed is measured by the distance $d_s$, measured in feet, a golf ball rolls when launched from a Stimp meter on a level (i.e. 0% grade) surface [4].

This paper will focus on numerical calculations of putt trajectories and analytical approximations to these trajectories. The accuracy of these approximations is intended to be better than ±1 inch, which amounts to a ±0.5 degree lateral alignment error at a distance of 10 feet. This figure of merit is based on two relevant facts. First, the diameter of the hole is 4.25 inches, and thus errors of order ±1 inch still place the ball within the hole. Second, Broadie's recent statistical analysis of putting [5] suggests the average PGA tour player hits putts with errors of order ±1.5 degrees. Thus, computational errors less than ±0.5 degree are three times smaller than what the average PGA tour player can achieve.

The equations of motion for a golf ball as it rolls across a putting green have been detailed by other authors, most recently by Penner [6]. The analysis in this paper is similar to that of Penner, but with a slightly different formalism that is thoroughly documented in Appendix A. An important generalization obtained from this new formalism is that the putting surface affects the putt trajectory through a single parameter, which is the product of the Stimp distance $d_s$ times the grade of the green, $\theta$. As an example, putts on a surface with a Stimp 8 speed and 1.5% grade follow the same trajectories as putts on a surface with Stimp 12 speed and 1% grade. This paper characterizes putting surfaces in terms this Stimp-grade product $d_s\theta$, measured in units of ft-%. The above examples are 12 ft-% putting surfaces.

Figure 1 displays an example of one such putt trajectory. The putt originates 10 feet from the hole on a 20 ft-% putting surface. Describing the initial angular position of the putt relative to the fall line by using the numbers on a face of a clock, with 12 o'clock directly uphill from the hole, this putt originates from the 8 o'clock position. The putt trajectory, shown in Fig. 1 as the solid black line, is that trajectory for which the ball crosses the center of the hole with a terminal speed which would have allowed the ball to roll 18 inches past the hole. This criterion is used for all calculations presented in this paper. The target line, shown in Fig. 1 as the dashed blue line, is drawn through the initial position of the ball and tangent to the initial trajectory of the putt. Fig. 1a is 10ft x 10ft field of view, which shows the entire putt trajectory. Fig. 1b is a 2ft x 2ft field of view in the vicinity of hole. All images of putt trajectories in this paper will have these two fields of view.

The traditional method of lining up a putt is to specify the target line in terms of its distance of closest approach to the hole, indicated in Fig. 1(b) as the red dot. The putt shown in Fig. 1 has a target line which aligns 5 inches left of the hole.

H.A. Templeton, the Target Point, and the Stimp Meter

This paper considers families of putts equidistant from the hole (i.e. originating on a circle centered on the hole). Figure 2 shows trajectories of a set of five putts, all originating 10 feet from the hole on a 20 ft-% putting surface. The five putts originate from the 7 o'clock thru 11 o'clock positions. The conventional means of aligning the target lines are indicated by red circles. From this perspective, these five putts seem to be unrelated.

Fig. 3 shows the same five putts, but with the target lines extended. Clearly, the target lines of these putts converge near to a point on the fall line, indicating this set of putt trajectories are actually closely related.

Fig. 4 shows the full family of 10 foot putt trajectories on this 20 ft-% putting surface. The target lines of all these putts converge in the vicinity of an average target point located on the fall line, indicated in Fig. 4(b) by a red dot. The algorithm used to calculate this average target point is detailed in Appendix B.

The concept of a target point was first described by H.A. Templeton in his book *Vector Putting*. Templeton's central thesis was that the target lines of all putts in the family of equidistant putts converge near to a single point, a target point. This is trivially true in the limit of a level green, as the target point is the center of the hole. Templeton proposed that as the grade of the green, speed of the green, and length of the putt increases, a target point persists but its position moves up the fall line, i.e. directly uphill from the hole. The trajectories shown in Figs. 5-9 are a visual example of this concept. These figures show families of putt trajectories as the putt length increases from 5 to 15 feet on a 20 ft-% putting surface. Templeton's target point is shown as the red dot in

Figs. 5(b) – 9(b), and it clearly moves up the fall line as the putt length increases. Also note that the region over which the target lines converge becomes larger as the putts get longer and the target point moves further up the fall line. Thus, one might argue that the concept of a target point becomes less relevant as the putt gets longer.

Templeton provided detailed tables in Appendix A of *Vector Putting*, listing the target point as a function of Stimp distance, grade, and putt length. Templeton was well aware that the target lines did not converge at a single point, but rather cross the fall line in a region centered on the target point. The tables in his Appendix A indicate the size of this region.

Fig. 10 compares the target points listed in Appendix A of *Vector Putting* for the case of a Stimp 6.5 green speed with the calculations generated using the formalism documented in Appendix A of this paper. Templeton's data is shown as open circles connected by dashed lines while the calculations of this paper are indicated as stars connected by solid lines. Note that Templeton's data consistently underestimates the position of the target point relative to the calculations of this paper. While Templeton mentions that the putt trajectories were calculated using a computer program, this program is not documented in his book. Additionally, Templeton provides almost no documentation in *Vector Putting* on the equations of motion used to calculate his target points. However, Templeton does discuss in detail the Stimp meter, which provides a clue as to why the two data sets displayed in Fig. 10 are not identical.

All putt trajectories are predicated on knowing the drag force on the golf ball as it rolls across the putting green. This drag force $F_d$, and its resulting acceleration $a_d$, is calculated from knowledge of the initial speed $v_s$ with which the ball is launched from the Stimp meter [3] and from the resulting Stimp distance $d_s$. Assuming the drag force is constant throughout the entire trajectory, conservation of energy yields the relation $\frac{1}{2}mv_s^2 = F_d d_s$, where $m$ is the mass of the ball. The resulting acceleration is

$a_d = \frac{F_d}{m} = \frac{v_s^2}{2d_s}$. Thus the putt trajectory depends on the square of the assumed value of $v_s$. In analyzing the Stimp meter Templeton calculates $v_s = 6.5$ ft/s, which is based on the calculation of a ball rolling down a simple inclined plane [7]. In fact, the Stimp meter is slightly more complicated than Templeton's assumptions. Holmes' more complete analysis of the Stimp meter [8] suggests 6.0 ft/s is more accurate, though this is also a calculation unsubstantiated by measurement. One can not really claim to accurately know $v_s$ until it is directly measured, and this author is unaware of any such published measurement. It is likely the true value of $v_s$ is slightly smaller than 6.0 ft/s, as Holmes' calculations do not address inelastic effects, such as the impact of the ball with the putting surface as it exits the Stimp meter and the effect of the dimples as it rolls down the incline. Never-the-less, all the calculations in this paper assume $v_s = 6.0$ ft/s, unless otherwise stated.

Templeton's value $v_s = 6.5$ ft/s is at least 10% too large. Because the equations of motion depend on the square of this speed, all of the target points in his Appendix A are too small by approximately 20%. Shown in Fig. 11 is a comparison of Templeton's data with the calculations of this paper assuming $v_s = 6.5$ ft/s. This comparison of the data is much more favorable than what is shown in Fig. 10. It strongly suggests that Templeton's computer program was based on equations of motion very similar to what is described in Appendix A of this paper.  Another inference from this comparison is that Templeton's central thesis has not gained wider appeal in the golfing community because the target points listed in Appendix A of his book are 20% too small.

Target Points On A Diamond Substructure

One problem with the concept of a target point is that the target lines for families of equidistant putts do not converge at a single point, but rather cross the fall line in a region centered on the target point.  As the length of the putt, speed of the green, and grade of the green increase, the size of the target point region increases.  This blurring would seem to degrade the usefulness of the concept.  It is interesting to ask if the concept of a target point can be recovered by considering subsets of putt trajectories equidistant from the hole.

Shown in Fig. 12 thru Fig. 23 is the same family of trajectories shown in Fig. 4, but organized into subsets that span a ±30 degree angular range.  For instance, Fig. 12 shows a subset of putts centered on 6 o'clock spanning the range from 5 o'clock thru 7 o'clock.  Note that each subset of putts has a unique, well defined, target point indicated by the red dot in Fig 12(b).  Thus, the concept of a target point remains robust when considering subsets of putts equidistant from the hole within an arc spanning ±30 degrees.

The ensemble of target points of all subsets of putts map out a diamond-shaped sub-structure, shown as the green shape above the hole in Figs. 12(b) – 23(b).  This target diamond is centered on Templeton's target point.  As the origin of the putt moves clockwise around the circle of equidistant putts, the corresponding target point moves counter-clockwise around the diamond sub-structure. Thus, the broadening of the target point along the fall line reported by Templeton is actually a manifestation of mapping the circle of equidistant putts into the diamond of target points. As is described at the end of this paper, this understanding forms the basis of a methodology which can be used on the course to increase the probability that the golfer chooses the correct target line.

The Universal Curve

Calculations of putt trajectories have been performed as a function of putt length for a wide range of Stimp-grade products. In each instance, the center position (measured relative to the center of the hole), height and width of the target diamond are determined. Shown in Fig. 24 are the center positions of the target diamond as a function of putt length, for putt lengths from 2.5-15 feet, for values of Stimp-grade product from 5-30 ft-% and with a terminal speed which would have allowed the ball to roll 18 inches past the hole.

Shown in Fig. 25 is the height (red line) and width (blue line) of the target diamond for the same conditions shown in Fig. 24. Note that the target diamond is very symmetric, as the width and height are approximately equal over most of the range of parameters. The black dashed line in Fig. 25 is the average of the height and width and will be used as the measure of the size scale of the target diamond. The height and width are always well within one inch of the average.

All the data represented in Fig. 24 and Fig. 25 can be approximated by a universal curve, shown in Fig. 26. The left axis indicates the distance to the center of the target diamond normalized by the parameter $\xi = d_s \theta / 10$, i.e. the Stimp-grade product divided by 10. The right axis is the size scale of the target diamond normalized by $\xi^2$. This universal curve is determined as follows. The curves of Fig. 24 are fit to the functional form $y(x) = d_s\theta\, U_c(x)$, where $y$ is the center position of the target diamond, $x$ is the length of the putt, and $U_c(x)$ is a fourth order polynomial constrained to go through the origin,

$$U_c(x) = c_1 x + c_2 x^2 + c_3 x^3 + c_4 x^4.$$

The curves of Fig. 25 were fit to the functional form $y(x) = (d_s\theta)^2\, U_d(x)$, where $y$ is the size scale of the target diamond, $x$ is the length of the putt, and $U_d(x)$ is a fourth order polynomial constrained to go through the origin,

$$U_d(x) = d_1 x + d_2 x^2 + d_3 x^3 + d_4 x^4.$$

The resulting curves $U_c$ and $U_d$ very nearly satisfy the proportionality $U_c(x) = a U_d(x)$. A linear regression is performed to determine the optimal proportionality constant, $a = \sum_x U_c U_d \Big/ \sum_x U_d^2$. The universal curve $U(x)$ is calculated as the average, $U(x) = [U_c(x) + a U_d(x)]/2$.

This mathematical formulation enables approximation of the center position of the target diamond as a function of putt length and Stimp-grade product through the relation $C(x, d_s\theta) = (d_s\theta)\, a_c\, U(x)$, where $a_c$ is a scale factor. The value $a_c$ is determined by least squares fit of $C(x, d_s\theta)$ to the data of Fig. 24. The resulting scaling is indicated by the left axis of Fig. 26, which represents $C(x, d_s\theta)$ normalized by $\xi = d_s\theta/10$. The comparison between the original data (i.e. Fig. 24) and the universal curve approximation is shown in Fig. 27. The universal curve approximation is shown as the black lines and the numerical calculations of Fig. 24 are shown as the open circles connected by dashed lines. The maximum difference between the numerical calculation and the universal curve approximation is less than one inch over the entire data set, and thus the universal curve is a reasonable approximation to the center position of the target diamond.

Similarly, the size scale (i.e. average of height and width) of the target diamond is approximated by the function $HW(x, d_s\theta) = (d_s\theta)^2 \, a_{hw} \, U(x)$, where $a_{hw}$ is a scale factor. The value $a_{hw}$ is determined by least squares fit of $HW(x, d_s\theta)$ to the data of Fig. 25. The scaling is indicated by the right axis of Fig. 26, which represents $HW(x, d_s\theta)$ normalized by $\xi^2$. The resulting comparison of the original data (i.e. the dashed black line in Fig. 25) and the universal curve approximation to the data is shown in Fig. 28. The universal curve approximation is shown as the black lines and the numerical calculations of Fig. 25 are shown as the open circles connected by dashed lines. Again, the maximum difference between the numerical calculation and the universal curve approximation is less than one inch over the entire data set, and thus the universal curve is also a reasonable approximation to the height and width of the target diamond.

The universal curve shown in Fig. 26 is a complete summary of the geometry of putting in the limit of a planar putting surface for putts with a terminal speed which would have allowed the ball to roll 18 inches past the hole. The universal curve can be calculated for any reasonable terminal speed; higher speeds move the curve down while lower speeds move the curve up.

If one has prior knowledge of the length of the putt, the speed of the green and the grade of the green, the target point can be determined by using this universal curve and an understanding of the target diamond. While competitors on the PGA tour might have access to this type of information prior to putting, most golfers will not. The next section describes how golfers can use this understanding of the geometry of putting to improve the probability that the correct target line is chosen.

Reading Putts: Reducing the Standard Error of the Mean of the Target Point

One of the great challenges facing every golfer is to correctly read the putt-at-hand, on the golf course, while playing in an event. The results presented above suggest a methodology for reading putts when the putting surface is reasonably approximated by a plane, which is often the case for putts less than 10-15 feet.

This methodology is based on well developed statistical analysis. Suppose one is making measurements of a noisy data source whose true underlying distribution is of mean $\mu$ and variance $\sigma^2$. Denote each measurement as $x_i$ and suppose one makes $N$ measurements. The goal is to determine the best estimates of the true distribution using the measured data. The best estimate of the mean is $\langle \mu \rangle = \frac{1}{N} \sum_{i=1}^{N} x_i$ and the best estimate of the variance is $\langle \sigma^2 \rangle = \frac{1}{N-1} \sum_{i=1}^{N} (x_i - <x>)^2$. One can ask how close will the estimate $\langle \mu \rangle$ be to the true value $\mu$. The answer is that the standard error of the estimate of the mean is $\pm \frac{\sigma}{\sqrt{N}}$. Thus, the more measurements that are made, the better one is able to estimate the true mean when measuring a noisy distribution.

This statistical analysis guides the methodology for reading putts advocated in this paper. The traditional way of lining up a putt is to consider only the putt-at-hand. In contrast, **<u>understanding that many putts share a common target point with the putt-at-hand strongly suggests that golfers should line up not just the putt-at-hand, but all the putts that share a common target point with the putt-at-hand</u>**.  In particular, the results presented in this paper suggest lining up many putts, all equidistant from the hole on an arc of order ±30 degrees relative to the putt-at-hand.  For each such putt, make a best estimate of the target line, and then determine the location of the target point by determining the place where all the target lines cross.  It is likely that after the first look not all of the target lines will cross at a point.  The golfer should iterate through this process, adjusting estimated target lines until a single target point is determined.

When assessing the reliability of this estimate of the target point and iterating towards a solution, it may be useful to understand the details of the target diamond, but in practice all that is necessary to know is that the target lines of related putts will cross at the target point.

The advantage of this methodology is that it requires no quantitative calculation, and thus no exact knowledge of the speed and slope of the green and length of the putt.  The golfer uses instinct, experience, and imagination to estimate the trajectory of a putt, just as has always been done.  However, by considering a family of putts all known to share the same target point, the golfer increases the probability of correctly choosing the correct target point, and thus the correct target line.

Summary

This paper explores the geometry of putting in the limit of a planar putting surface, an approximation which is often accurate within 10-15 feet of the hole.  Putts are organized into subsets of putts equidistant from the hole on an arc spanning ±30 degrees. It is shown that each of these subsets of putts share a common target point. Moving in a circle around the perimeter of equidistant putts, the target point maps out a diamond-shaped structure centered on the fall line directly above the hole. As the length of the putt, speed of the green, and/or grade of the green increase, the position of the target diamond moves further up the fall line and the size of the target diamond increases. It is shown that a single universal curve can be used to determine the dimensions of the target diamond for a putt of any length on a planar putting surface with any speed and grade.

While these computational results may seem esoteric, in fact they suggest a very simple technique for reading putts that can improve the probability of properly choosing the proper target line. Conventionally, golfers only line up the putt-at-hand. The results presented in this paper suggest lining up many putts, all equidistant from the hole on an arc of order ±30 degrees relative to the putt-at-hand and using all the information from this exercise to estimate the common target point. By considering a family of putts all known to share the same target point, the golfer increases the probability of correctly identifying the target point, and thus the correct target line.

Acknowledgements: The author is grateful for fun and insightful conversations with Lars Bildsten, Mark Broadie, Shawn Cox, Will Goetzmann, Michael Hebron, Jeff Kimble, Phil Rogers, and Grant Waite.

Appendix A:  The Equations of Motion

The equations of motion of putting have been previously summarized by Penner [6]. The following analysis is similar to that of Penner, but with a slightly different formalism. As will be shown, this different formalism allows for some useful generalizations about putt trajectories.

The approximation made throughout this paper is that the ground near to the hole approximates a tilted plane. The analysis starts by considering a flat surface normal to the gravitational force, at the center of which is fixed a Cartesian coordinate system $\hat{x}$, $\hat{y}$, and $\hat{z}$, in which the $\hat{z}$ axis is perpendicular to the plane and parallel to the gravitational force. The plane is then tilted by rotating about the $\hat{x}$ axis through an angle $\theta$ (i.e. rotating the $\hat{y}$ axis towards the $\hat{z}$ axis). This suggests a new coordinate system $\hat{x}'$, $\hat{y}'$, $\hat{z}'$ in the inclined plane, defined by the transformation

$$\begin{pmatrix} x' \\ y' \\ z' \end{pmatrix} = \begin{pmatrix} 1 & 0 & 0 \\ 0 & \cos\theta & \sin\theta \\ 0 & -\sin\theta & \cos\theta \end{pmatrix} \begin{pmatrix} x \\ y \\ z \end{pmatrix},$$

where the uphill direction is $\hat{y}'$.

Because the equations of motion are linear, we can discuss the forces on the ball as two independent problems. These are the gravitational and drag forces.

The gravitational force on the ball acts through the center of mass and is equal to $\vec{F}_g = -mg\,\hat{z}$. It is convenient to divide this force into a component normal to the plane and a component in the plane, $\vec{F}_g = -mg(\hat{y}'\sin\theta + \hat{z}'\cos\theta)$. For a ball rolling on a surface, all forces through the center of mass and parallel to the plane must be accompanied by a force of static friction, $\vec{F}_s$. This force is conceived as originating at the point of contact between the ball and the plane and is of magnitude and orientation to keep the ball rolling without sliding. For a solid sphere of radius $R$ and inertial moment $I = \frac{2}{5}mR^2$, the drag force is $\vec{F}_s = \frac{2}{7}mg\,\hat{y}'\sin\theta$. The net force on the center of mass of the sphere in the plane of motion is $(\vec{F}_g + \vec{F}_s)\cdot\hat{y}' = -\frac{5}{7}mg\sin\theta$.

The drag force on the golf ball was reviewed in detail by Penner. In summary, the ball feels a drag force which is assumed constant over the entire trajectory of motion. The origin of this force involves the deformation of the grass under the ball. The force of gravity normal to the surface causes the ball to deform the surface on which it sits. As the ball rolls, it continuously deforms the surface under it. The drag force comes about because the surface does not immediately restore its shape, and so as the ball moves the back of the ball is not in contact with the deformed surface. As there is pressure only on the front surface of the ball, both the normal force and the drag force acts on the ball at a point of contact forward of the center of mass of the ball and in the direction of motion by the distance $\rho$.

The drag force is directly related to the Stimp distance, $d_s$, a standard defined by the United States Golf Association (USGA) for measuring the speed of the green [3]. The Stimp distance, measured in feet, is the distance a golf ball travels on a level surface when launched from a Stimp meter. The Stimp meter is a 30-inch long, V-grooved, inclined plane held at an angle of 20 degrees relative to the putting surface down. It has been reasoned that the Stimp meter launches the golf ball with an initial speed $v_s = 1.83$ m/s (6.0 ft/s) [8], thought it is not clear that its direct measurement on real putting surfaces has ever been published [9].

One can relate the drag force $\vec{F}_d$ to the Stimp distance $d_s$ by assuming $\vec{F}_d$ is constant over the entire trajectory and by considering the work done by the drag force to stop the ball, $F_d d_s = \dfrac{m v_s^2}{2}$. Throughout this text, the ratio of the drag force to the gravitational force, $\dfrac{F_d}{mg} = \dfrac{v_s^2}{2 d_s g}$, is often referenced.

The direction of the drag force is opposite to the direction of travel of the ball. The direction of travel is defined by the angle $\phi$, which is measured as a rotation from the $\hat{x}'$ axis towards the $\hat{y}'$ axis. Thus, the velocity vector is given as $\vec{v} = v(\hat{x}' \cos\phi + \hat{y}' \sin\phi)$ and the drag vector is $\vec{F}_d = -F_d(\hat{x}' \cos\phi + \hat{y}' \sin\phi)$. Also note that when graphing trajectories, $y'(x')$, the angle $\phi$ is defined as $\tan\phi = \dfrac{dy'}{dx'}$.

The calculations in this paper assume the ball rolls without slipping. The no-slip condition is maintained by balancing the various torques acting on the ball. The normal force, $\vec{F}_g \cdot \hat{z}' = mg\cos\theta$, provides a torque of magnitude $\rho mg\cos\theta$ that reduces the rotational speed of the ball. The drag force provides a net torque of magnitude $RF_d$, where $R$ is the radius of the ball, that increases the rotational speed of the ball. The total torque on the ball is $I\dot{\omega} = RF_d - \rho mg\cos\theta$, where $\omega$ is the rotational speed of the ball. The drag force on the ball yields the linear acceleration $m\dfrac{dv}{dt} = -F_d$. The no-slip condition is $\dfrac{dv}{dt} = R\dot{\omega}$, from which we obtain $\dfrac{F_d}{mg} = \dfrac{5}{7}\dfrac{\rho}{R}\cos\theta$. This serves to define the distance $\rho$ but otherwise has no effect on the motion of the ball.

Combining the drag force and the gravitational force, one obtains the equations of motion of the ball:

$$m\frac{d^2x'}{dt^2} = -F_d \cos\phi$$

$$m\frac{d^2y'}{dt^2} = -F_d \sin\phi - \frac{5}{7}mg \sin\theta$$

Factoring out the drag force,

$$\frac{mg}{F_d}\frac{d^2x'}{dt^2} = -g\cos\phi$$

$$\frac{mg}{F_d}\frac{d^2y'}{dt^2} = -g\sin\phi - \frac{5}{7}\frac{mg}{F_d}g\sin\theta$$

and using the expression relating $F_d$ to $d_s$, one obtains:

$$\frac{mg}{F_d}\frac{d^2x'}{dt^2} = -g\cos\phi$$

$$\frac{mg}{F_d}\frac{d^2y'}{dt^2} = -g\sin\phi - \frac{5}{7}\frac{2gd_s}{v_s^2}g\sin\theta$$

Rescaling time such that $\tau = \sqrt{\frac{F_d}{mg}}\,t = \sqrt{\frac{v_0^2}{2gd_s}}\,t$, one obtains the dimensionless equations

$$\frac{1}{g}\frac{d^2x'}{d\tau^2} = -\cos\phi$$

$$\frac{1}{g}\frac{d^2y'}{d\tau^2} = -\left(\frac{10}{7}\frac{g\,d_s\sin\theta}{v_s^2}\right) - \sin\phi.$$

These equations depend only on the scale factor $d_s \sin\theta$. For almost any reasonable pitch of a green, one can approximate $\sin\theta \approx \theta$, and thus the equations of motion depend only on the parameter, $d_s\theta$. Physically, this means that increasing the tilt of the green yields trajectories equivalent to a proportional increase in the Stimp value of the green, as characterized by $d_s$.

As defined above, the normalized time $\tau$ is related to real time through the factor $\tau = \sqrt{\dfrac{F_d}{mg}}\, t$, and thus the normalized velocity is related to the real velocity through the expression $\bar{v} = \dfrac{dx}{d\tau} = \sqrt{\dfrac{mg}{F_d}}\dfrac{dx}{dt} = \sqrt{\dfrac{mg}{F_d}}\, v$. The analysis below will calculate trajectories by assuming the ball enters the hole in the center of the hole traveling at a defined speed, $v_f$. Traveling at this speed on a level surface, the ball would come to rest a distance $d_p = \dfrac{m v_f^2}{2 F_d}$ past the hole. The normalized speed is related to the actual speed through the relation described above $\bar{v}_f = \sqrt{\dfrac{mg}{F_d}}\, v_f$. Relating this back to $d_p$, one obtains $\bar{v}_f = \sqrt{2 g d_f}$ for the normalized terminal velocity on a level green. On a tilted green one must correct for the gravitational acceleration, yielding the expression

$$\bar{v}_f = \sqrt{2 g d_f \left(1 + \dfrac{10}{7}\dfrac{g d_s \sin\theta \sin\phi}{v_s^2}\right)}.$$

Putt trajectories are obtained as solutions to the differential equations

$$\dfrac{1}{g}\dfrac{d^2 x'}{d\tau^2} = -\cos\phi$$

$$\dfrac{1}{g}\dfrac{d^2 y'}{d\tau^2} = -\dfrac{10}{7}\dfrac{g d_s}{v_s^2}\sin\theta - \sin\phi$$

subject to the terminal conditions $x_f = y_f = 0$ and $\bar{v}_f$ as defined above. This is done by starting at the final condition and letting time run backwards.

Appendix B: Calculating the Target Point

A primary result of this paper is that the target lines of families of related putts all share a common target point. This is an approximation, as the lines do not all cross at a single point. In fact, they all approach near to a single point. In this appendix we calculate the point nearest to a family of straight lines.

Assume a line through the point $(x_n, y_n)$ of slope $a_n$ given by the expression $\vec{R}_n(\varepsilon) = \langle x_n + \varepsilon, y_n + a_n \varepsilon \rangle$. The square of the distance from a point on the line to the point $(x_0, y_0)$ is $d_n(\varepsilon)^2 = (x_n + \varepsilon - x_0)^2 + (y_n + a_n \varepsilon - y_0)^2$. Minimizing this parameter with respect to $\varepsilon$, $\dfrac{\partial d_n^2}{\partial \varepsilon} = 0$, determines the point on the line which is closest to the point $(x_0, y_0)$. The solution occurs for the value $\varepsilon = -\dfrac{[a_n(y_n - y_0) + (x_n - x_0)]}{(1 + a_n^2)}$ and has the value $d_{\min,n}^2 = \dfrac{[(y_n - y_0) - a_n(x_n - x_0)]^2}{(1 + a_n^2)}$.

Now consider a family of lines, each line described in terms of an initial point $(x_n, y_n)$ and slope $a_n$. Assume one would like to find the point $(x_0, y_0)$ which is the point closest to all lines in the family of lines, as measured by minimizing the sum of the square of the minimum distances to the reference point,

$$d^2 = \sum_n \frac{[(y_n - y_0) - a_n(x_n - x_0)]^2}{(1 + a_n^2)}.$$

Minimizing this expression yields the two equations

$$\frac{\partial d^2}{\partial y_0} = -2 \sum_n \frac{1}{(1 + a_n^2)} [(y_n - y_0) - a_n(x_n - x_0)] = 0$$

$$\frac{\partial d^2}{\partial x_0} = 2\sum_n \frac{a_n}{(1+a_n^2)}[(y_n - y_0) - a_n(x_n - x_0)] = 0$$

which can be rearranged into the form

$$\sum_n \frac{1}{(1+a_n^2)} y_n - \sum_n \frac{a_n}{(1+a_n^2)} x_n = \left(\sum_n \frac{1}{(1+a_n^2)}\right) y_0 - \left(\sum_n \frac{a_n}{(1+a_n^2)}\right) x_0$$

$$\sum_n \frac{a_n}{(1+a_n^2)} y_n - \sum_n \frac{a_n^2}{(1+a_n^2)} x_n = \left(\sum_n \frac{a_n}{(1+a_n^2)}\right) y_0 - \left(\sum_n \frac{a_n^2}{(1+a_n^2)}\right) x_0$$

It is convenient to rewrite the entire expression as a matrix

$$\begin{pmatrix} \langle y \rangle - \langle ax \rangle \\ \langle ay \rangle - \langle a^2 x \rangle \end{pmatrix} = \begin{pmatrix} \langle 1 \rangle & -\langle a \rangle \\ \langle a \rangle & -\langle a^2 \rangle \end{pmatrix} \begin{pmatrix} y_0 \\ x_0 \end{pmatrix},$$

where the generic notation $\langle \xi \rangle = \frac{1}{N} \sum_n \frac{1}{(1+a_n^2)} \xi_n$ has been used to simplify the expressions. This matrix equation is readily inverted, yielding the solution for the point $(x_0, y_0)$,

$$\begin{pmatrix} y_0 \\ x_0 \end{pmatrix} = \frac{1}{(\langle a \rangle^2 - \langle 1 \rangle \langle a^2 \rangle)} \begin{pmatrix} -\langle a^2 \rangle & \langle a \rangle \\ -\langle a \rangle & \langle 1 \rangle \end{pmatrix} \begin{pmatrix} \langle y \rangle - \langle ax \rangle \\ \langle ay \rangle - \langle a^2 x \rangle \end{pmatrix}.$$

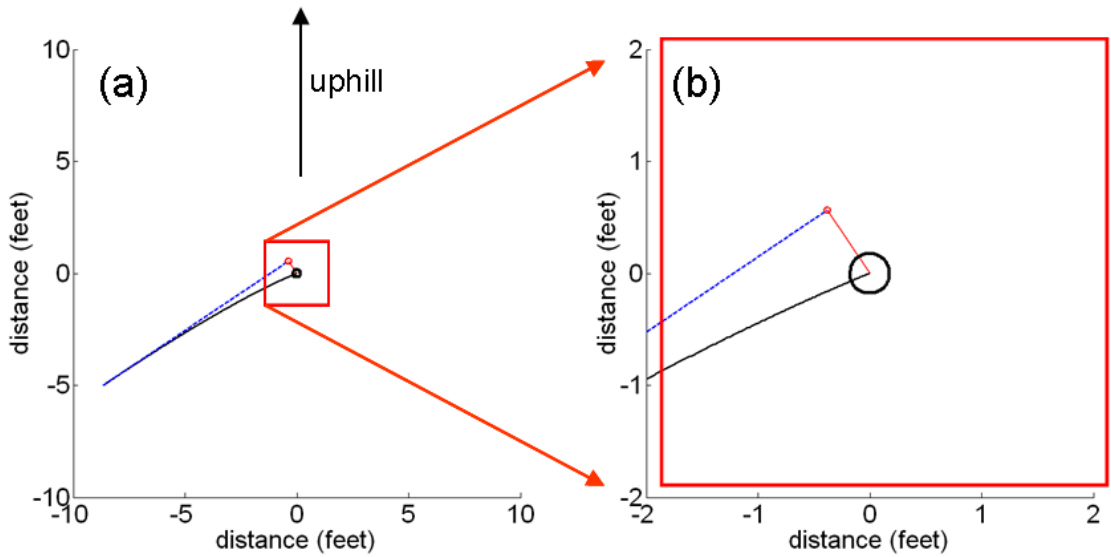

Figure 1: A 10 foot putt on a Stimp-grade 20 ft-% putting surface originating at the 8-o'clock position relative to the fall line.  The putt trajectory is shown as the solid black line.  The target line is shown as the dashed blue line.  Fig 1(a) is a 10ft x 10ft field of view showing the entire putt trajectory.  Fig. 1(b) is a 2ft x 2 ft field of view showing detail near to the hole.  The hole is shown as the back circle.  The putt trajectory is calculated for a putt crossing the center of the hole with a speed which would have carried it 18 inches beyond the hole.  The conventional means of aligning the target line is indicated by red circle, which is the nearest approach of the target line to the hole..

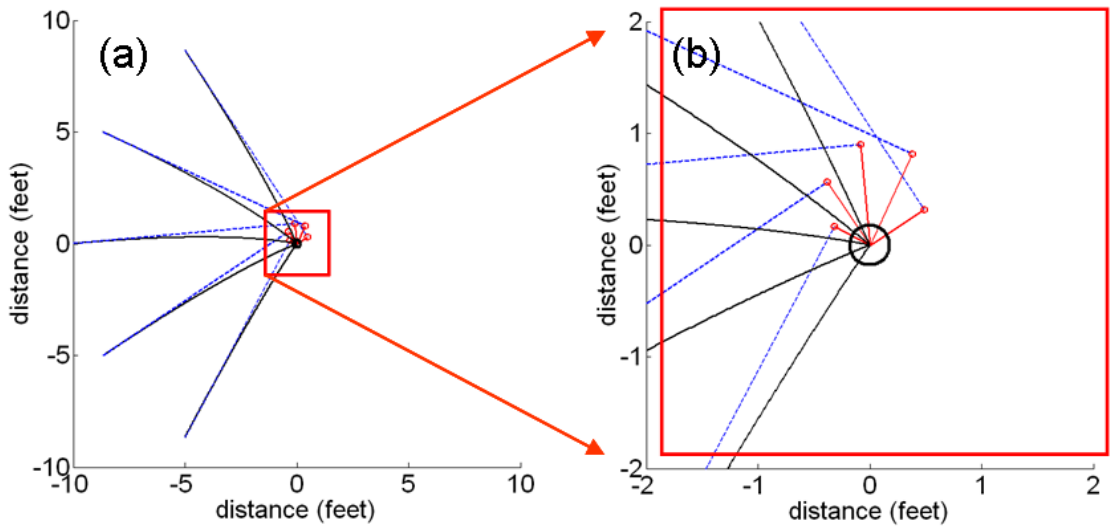

Figure 2:  Putt trajectories for five putts originating 10 feet from the hole on a 20 ft-% putting surface.  The five putts originate from the 7 o'clock thru 11 o'clock positions.  The conventional means of aligning the target lines are indicated by red circles.  From this perspective, all of these five putts seem unrelated.

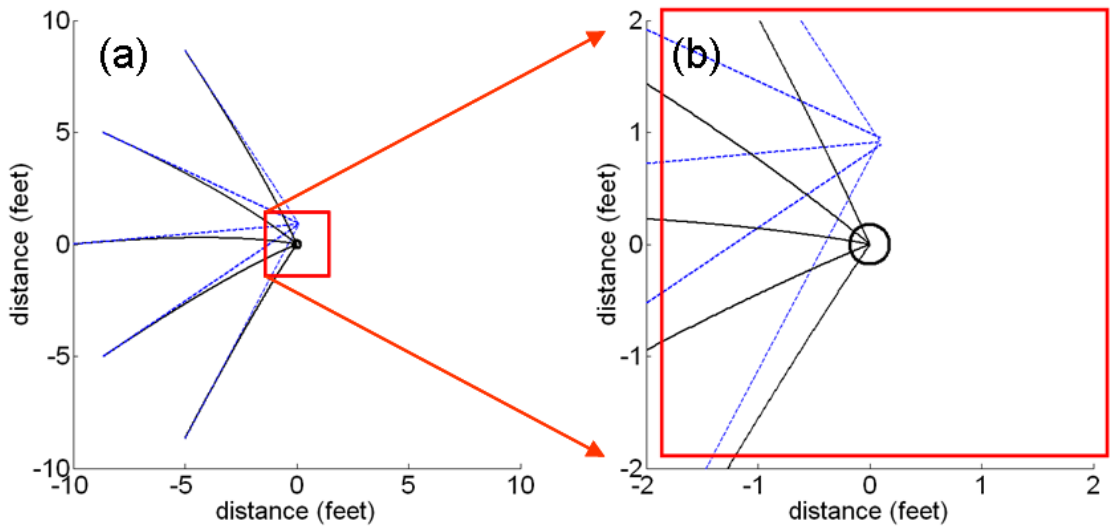

Fig. 3. The same five putts shown in Fig. 2, but with the target lines extended. The target lines of these putts converge near to a point on the fall line, indicating that these putts are actually closely related.

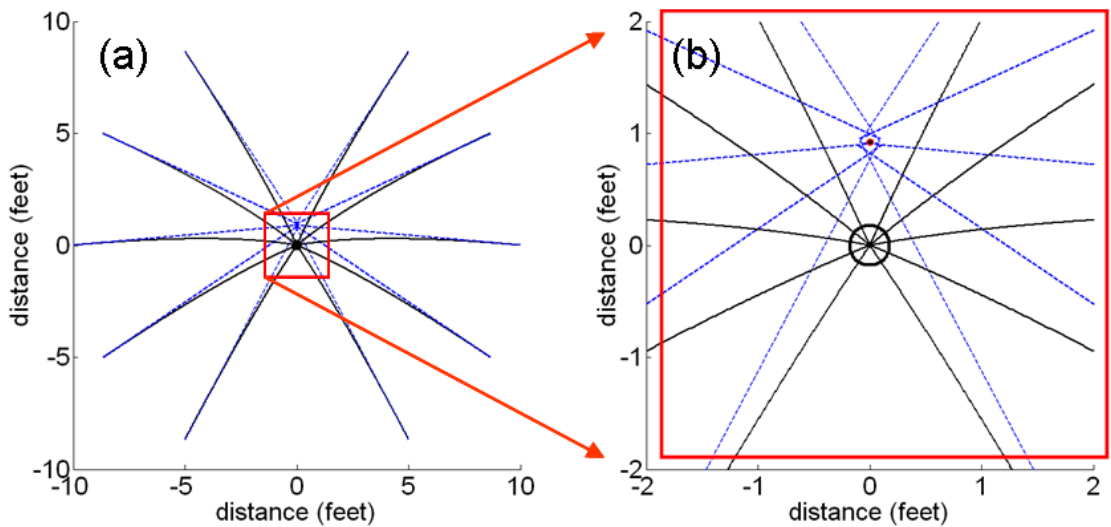

Fig. 4. The full family of 10 foot putt trajectories. The target lines of all these putt trajectories converge in the vicinity of an "average target point" located on the fall line. The position of this average target point is indicated in (b) by a red dot and corresponds to Templeton's target point.

\

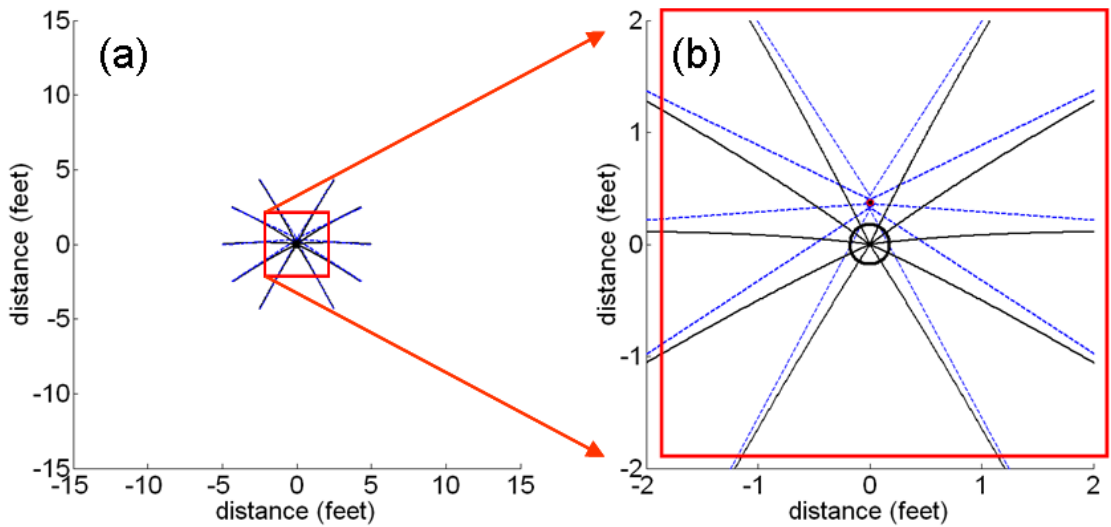

Fig. 5. The family of 5 foot putt trajectories on a Stimp-grade 20 ft-% putting surface. All trajectories cross the center of the hole with a speed which would have carried it 18 inches beyond the hole. The red dot in Fig. 5(b) indicates Templeton's target point.

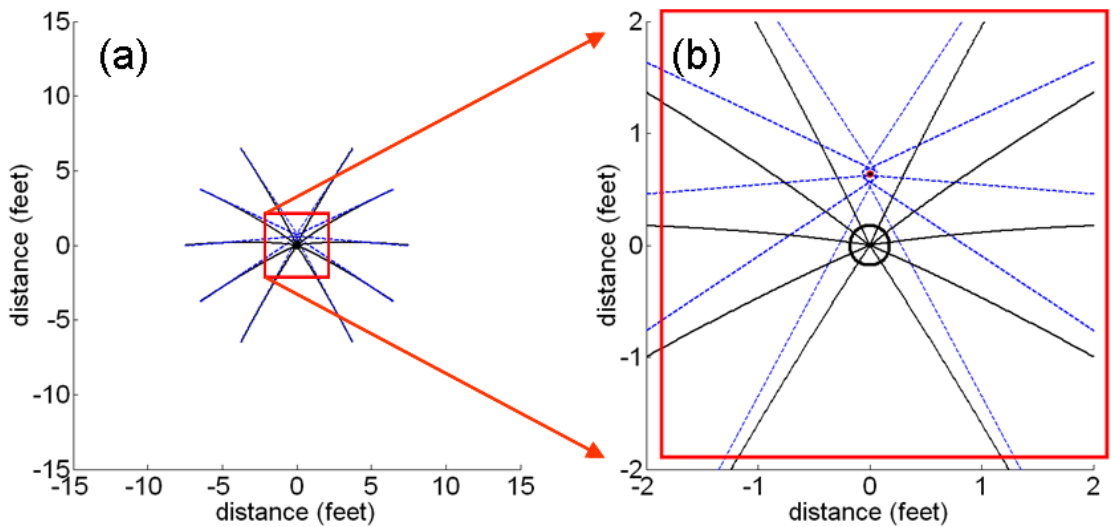

Fig. 6. The family of 7.5 foot putt trajectories on a Stimp-grade 20 ft-% putting surface. All trajectories cross the center of the hole with a speed which would have carried it 18 inches beyond the hole. The red dot in Fig. 6(b) indicates Templeton's target point.

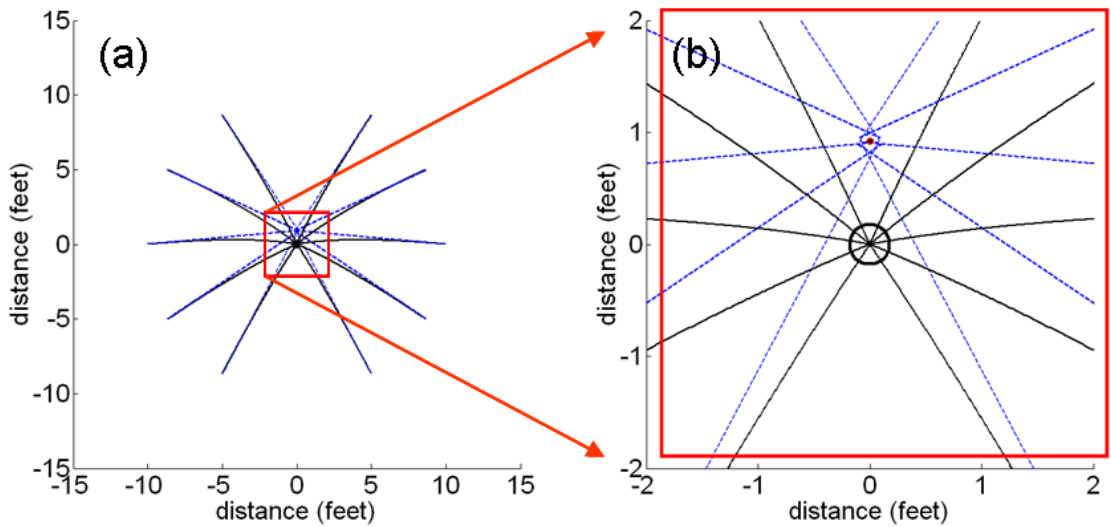

Fig. 7. The family of 10 foot putt trajectories on a Stimp-grade 20 ft-% putting surface. All trajectories cross the center of the hole with a speed which would have carried it 18 inches beyond the hole. The red dot in Fig. 7(b) indicates Templeton's target point.

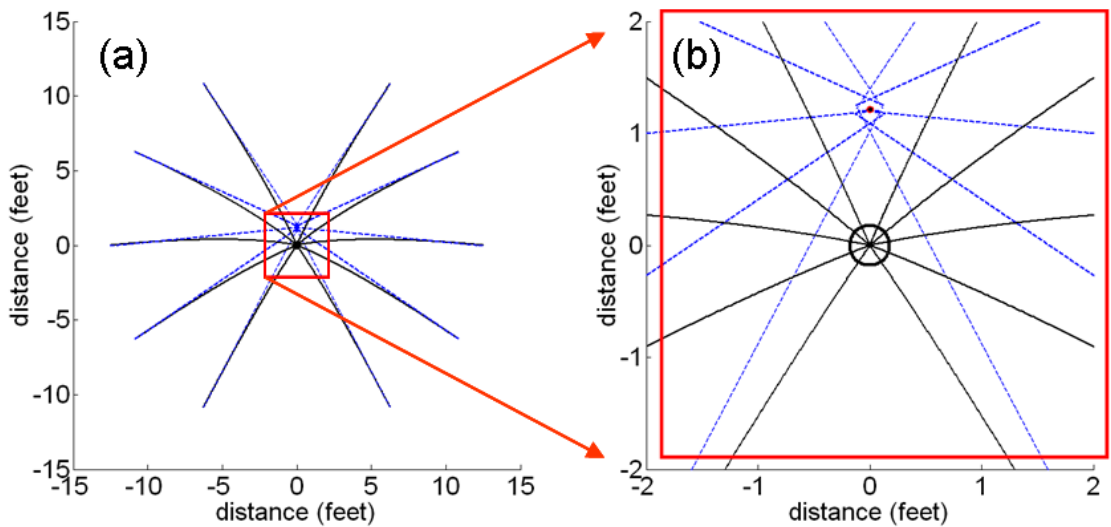

Fig. 8. The family of 12.5 foot putt trajectories on a Stimp-grade 20 ft-% putting surface. All trajectories cross the center of the hole with a speed which would have carried it 18 inches beyond the hole. The red dot in Fig. 8(b) indicates Templeton's target point.

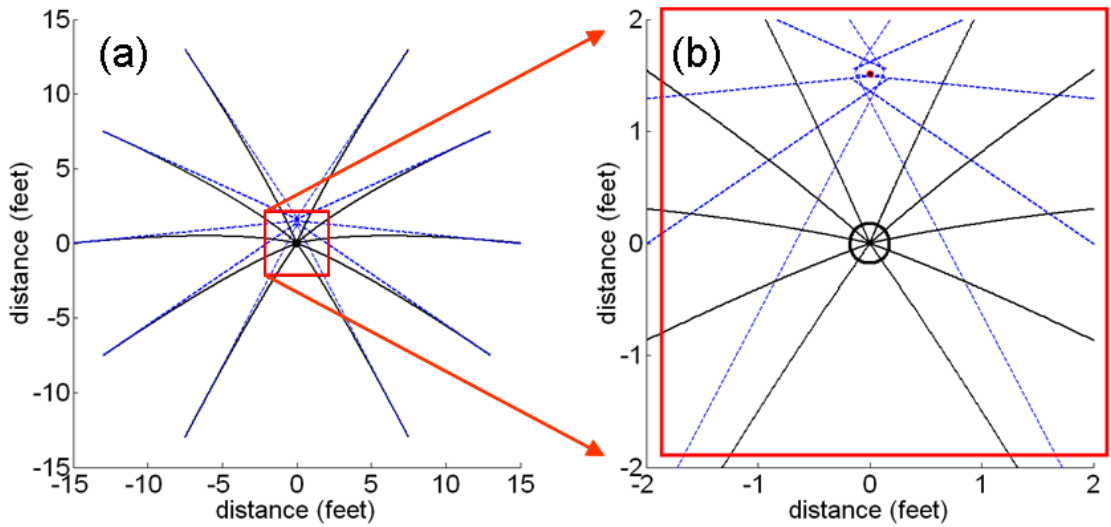

Fig. 9. The family of 15 foot putt trajectories on a Stimp-grade 20 ft-% putting surface. All trajectories cross the center of the hole with a speed which would have carried it 18 inches beyond the hole. The red dot in Fig. 9(b) indicates Templeton's target point.

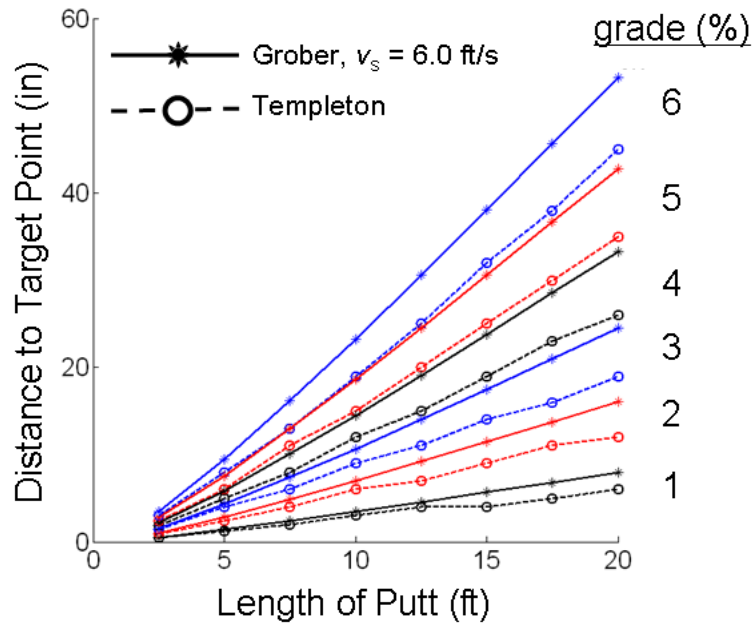

Fig. 10 compares the target points listed in Appendix A of *Vector Putting* for the case of a Stimp 6.5 green speed and grades from 1-6 % with the calculations generated using the formalism documented in Appendix A of this paper. The calculations assume the Stimp meter launches the ball with $v_s = 6.0$ ft/s, which sets the scale for the drag force. Note that Templeton consistently underestimates the distance to the target point relative to the calculations of this paper.

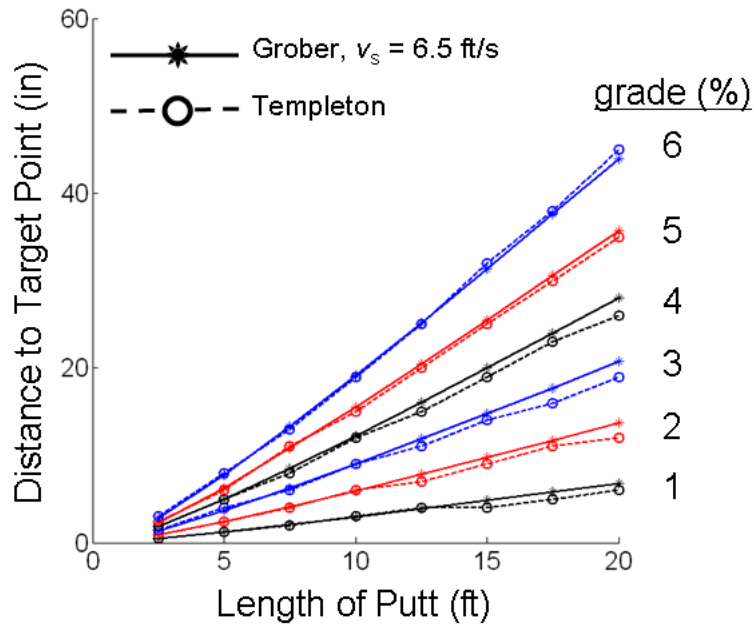

Fig. 11. The same comparison as in Fig. 10 but for calculations which assume $v_s = 6.5$ ft/s. Note that these calculations reproduce Templeton's results, suggesting Templeton's calculations were correct but he incorrectly correlated Stimp speed with drag force. As a result, the numbers in Appendix A of his book are all too small by approximately 20%.

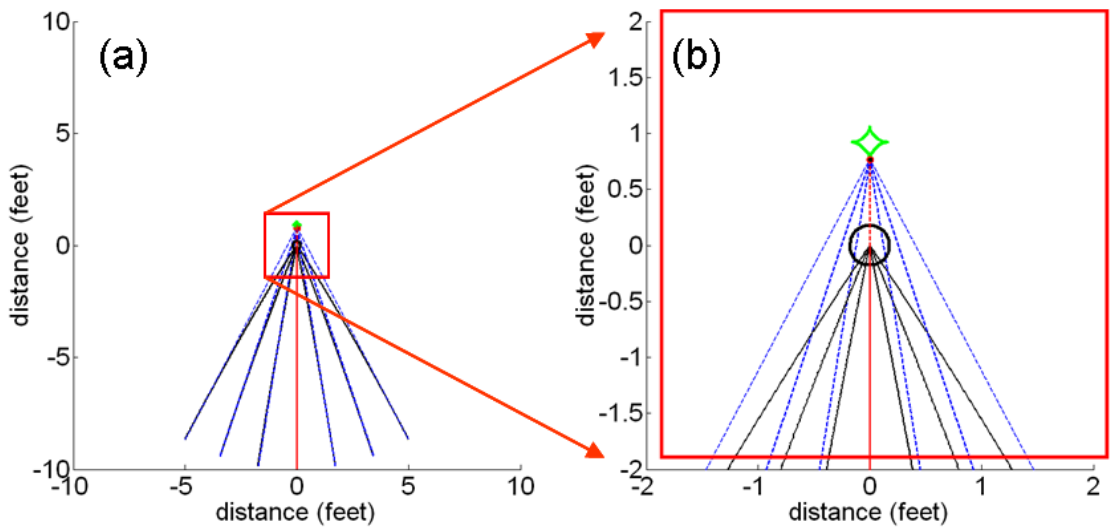

Fig. 12. Putt trajectories for 10 foot putts on a Stimp-grade 20 ft-% surface originating in a range of angles centered on the 6 o'clock position and spanning the range from 5 o'clock to 7 o'clock (i.e. a ±30 degree range of initial positions). As can be seen in Fig. 12(b), the target lines all cross very near to one another, sufficiently close that we can call it a target point. The target point is indicated by the red dot. The green diamond shaped structure above the hole is the ensemble of target points for all 10 foot putts on this putting surface, as can be seen in Figs. 12-23.

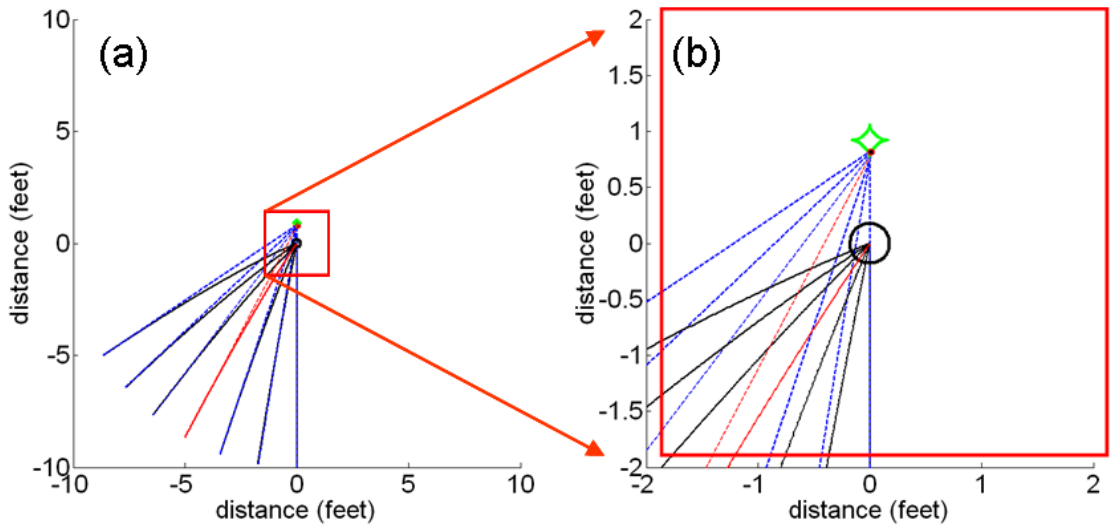

Fig. 13. Putt trajectories for 10 foot putts on a stimp-grade 20 ft-% surface originating in a range of angles centered on the 7 o'clock position and spanning the range from 6 o'clock to 8 o'clock. As can be seen in Fig. 13(b), the target lines all cross very near to one another, sufficiently close that we can call it a target point. The target point is indicated by the red dot. The green diamond shaped structure above the hole is the ensemble of target points for all 10 foot putts on this putting surface.

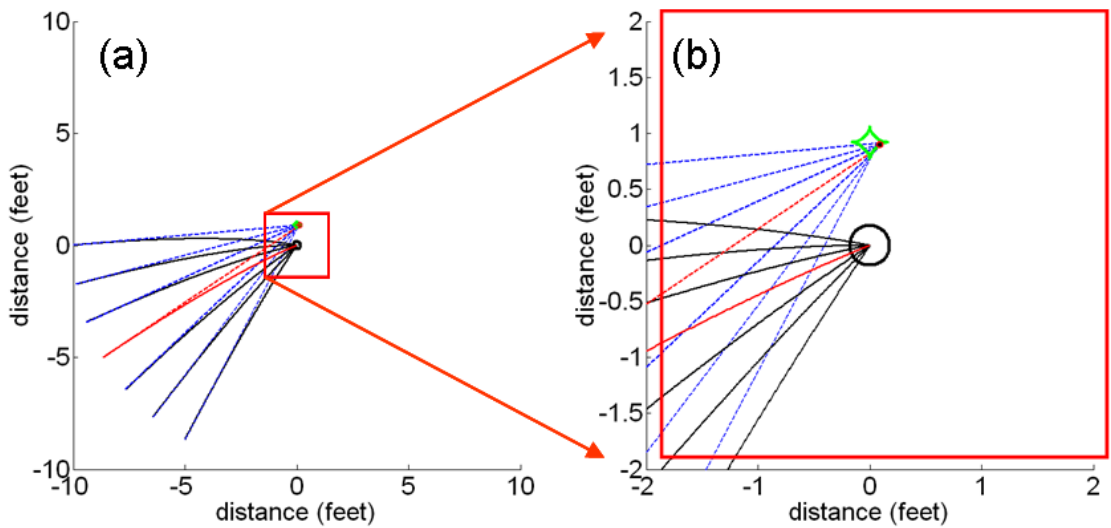

Fig. 14. Putt trajectories for 10 foot putts on a stimp-grade 20 ft-% surface originating in a range of angles centered on the 8 o'clock position and spanning the range from 7 o'clock to 9 o'clock. As can be seen in Fig. 14(b), the target lines all cross very near to one another, sufficiently close that we can call it a target point. The target point is indicated by the red dot. The green diamond shaped structure above the hole is the ensemble of target points for all 10 foot putts on this putting surface.

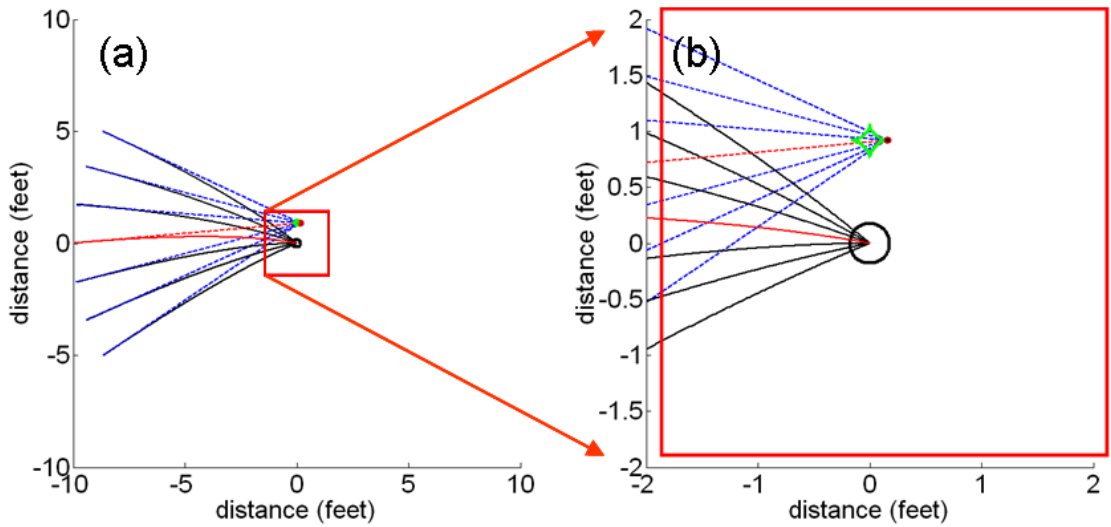

Fig. 15. Putt trajectories for 10 foot putts on a stimp-grade 20 ft-% surface originating in a range of angles centered on the 9 o'clock position and spanning the range from 8 o'clock to 10 o'clock. As can be seen in Fig. 15(b), the target lines all cross very near to one another, sufficiently close that we can call it a target point. The target point is indicated by the red dot. The green diamond shaped structure above the hole is the ensemble of target points for all 10 foot putts on this putting surface.

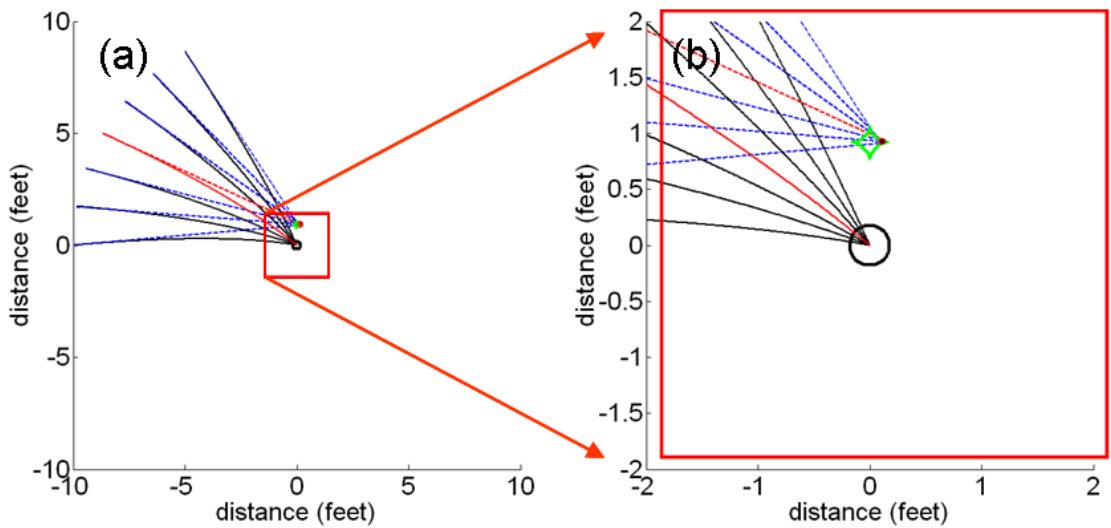

Fig. 16. Putt trajectories for 10 foot putts on a stimp-grade 20 ft-% surface originating in a range of angles centered on the 10 o'clock position and spanning the range from 9 o'clock to 11 o'clock. As can be seen in Fig. 16(b), the target lines all cross very near to one another, sufficiently close that we can call it a target point. The target point is indicated by the red dot. The green diamond shaped structure above the hole is the ensemble of target points for all 10 foot putts on this putting surface.

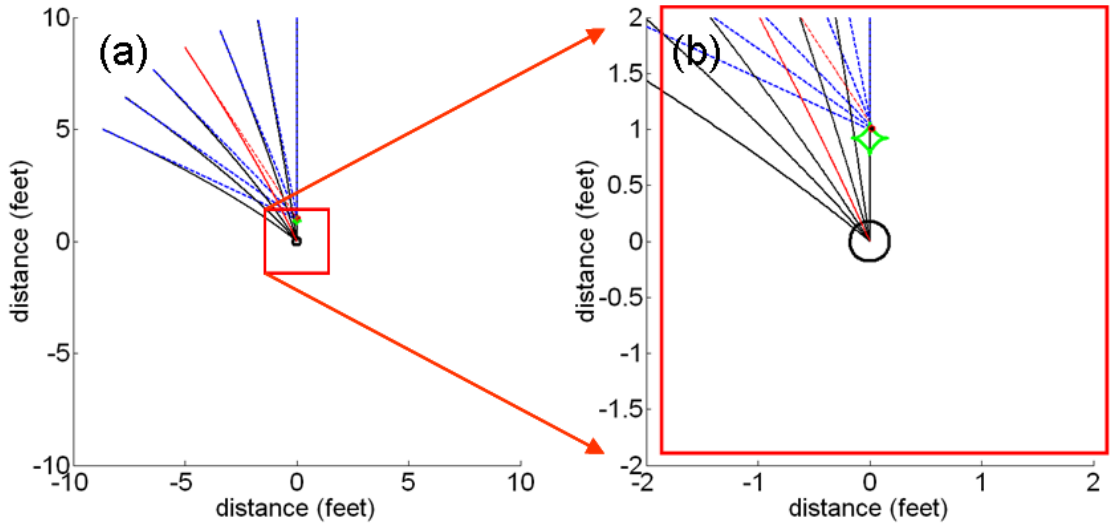

Fig. 17. Putt trajectories for 10 foot putts on a stimp-grade 20 ft-% surface originating in a range of angles centered on the 11 o'clock position and spanning the range from 10 o'clock to 12 o'clock. As can be seen in Fig. 17(b), the target lines all cross very near to one another, sufficiently close that we can call it a target point. The target point is indicated by the red dot. The green diamond shaped structure above the hole is the ensemble of target points for all 10 foot putts on this putting surface.

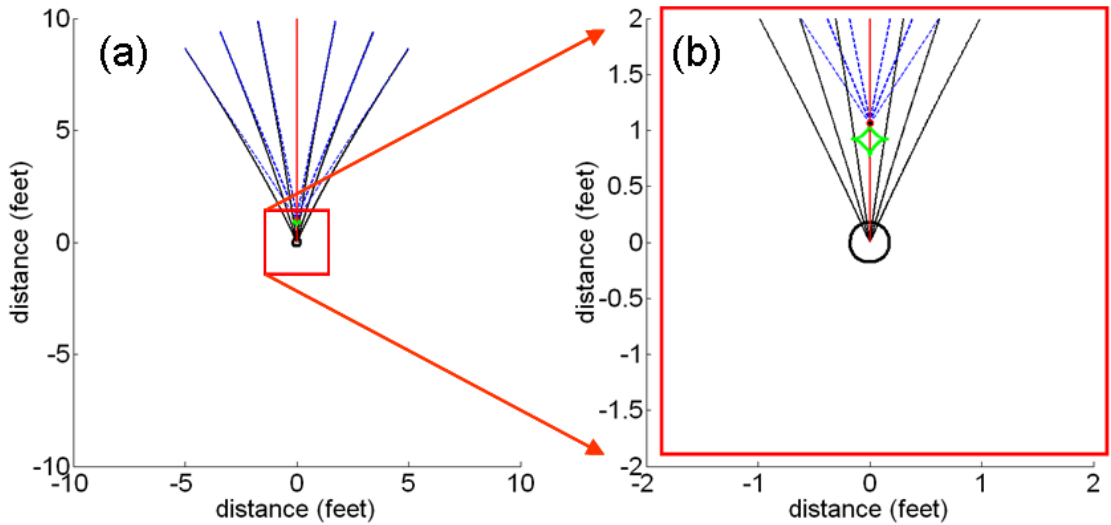

Fig. 18. Putt trajectories for 10 foot putts on a stimp-grade 20 ft-% surface originating in a range of angles centered on the 12 o'clock position and spanning the range from 11 o'clock to 1 o'clock. As can be seen in Fig. 18(b), the target lines all cross very near to one another, sufficiently close that we can call it a target point. The target point is indicated by the red dot. The green diamond shaped structure above the hole is the ensemble of target points for all 10 foot putts on this putting surface.

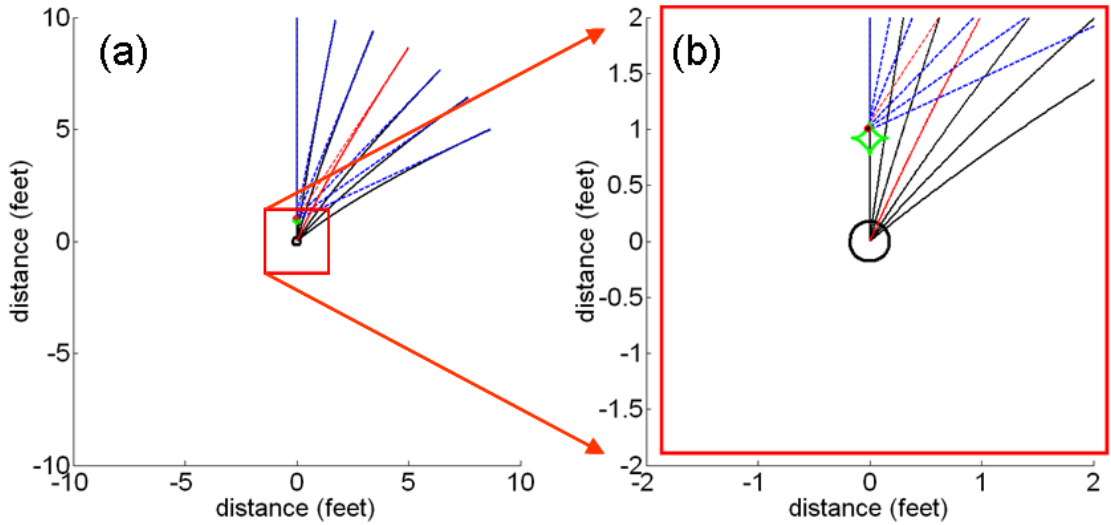

Fig. 19. Putt trajectories for 10 foot putts on a stimp-grade 20 ft-% surface originating in a range of angles centered on the 1 o'clock position and spanning the range from 12 o'clock to 2 o'clock. As can be seen in Fig. 19(b), the target lines all cross very near to one another, sufficiently close that we can call it a target point. The target point is indicated by the red dot. The green diamond shaped structure above the hole is the ensemble of target points for all 10 foot putts on this putting surface.

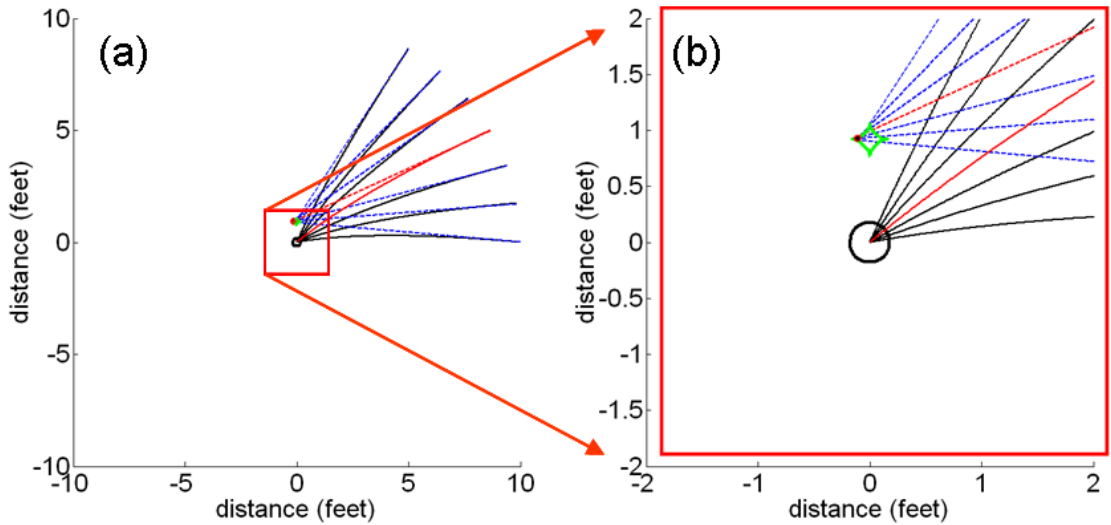

Fig. 20. Putt trajectories for 10 foot putts on a stimp-grade 20 ft-% surface originating in a range of angles centered on the 2 o'clock position and spanning the range from 1 o'clock to 3 o'clock. As can be seen in Fig. 20(b), the target lines all cross very near to one another, sufficiently close that we can call it a target point. The target point is indicated by the red dot. The green diamond shaped structure above the hole is the ensemble of target points for all 10 foot putts on this putting surface.

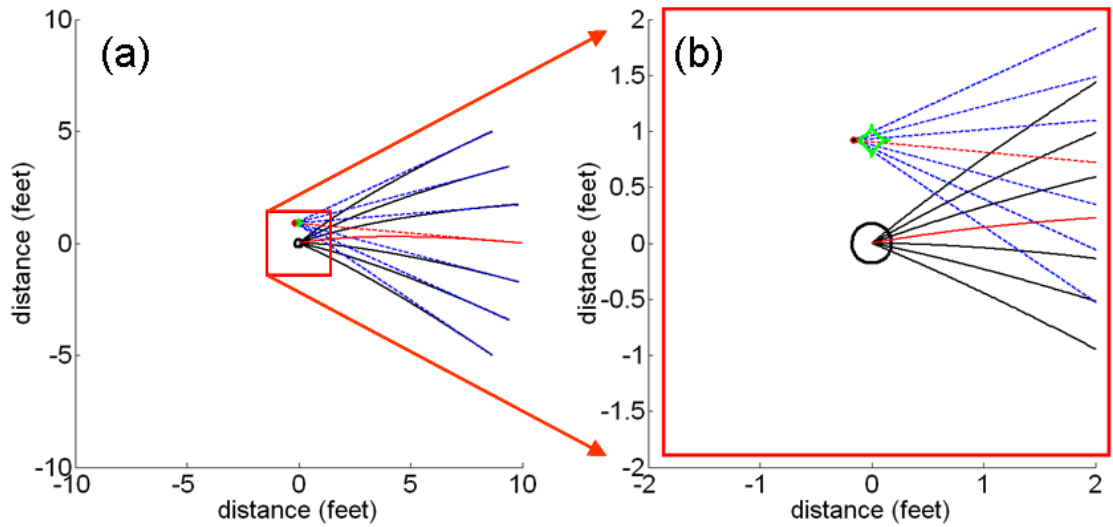

Fig. 21. Putt trajectories for 10 foot putts on a stimp-grade 20 ft-% surface originating in a range of angles centered on the 3 o'clock position and spanning the range from 2 o'clock to 4 o'clock. As can be seen in Fig. 21(b), the target lines all cross very near to one another, sufficiently close that we can call it a target point. The target point is indicated by the red dot. The green diamond shaped structure above the hole is the ensemble of target points for all 10 foot putts on this putting surface.

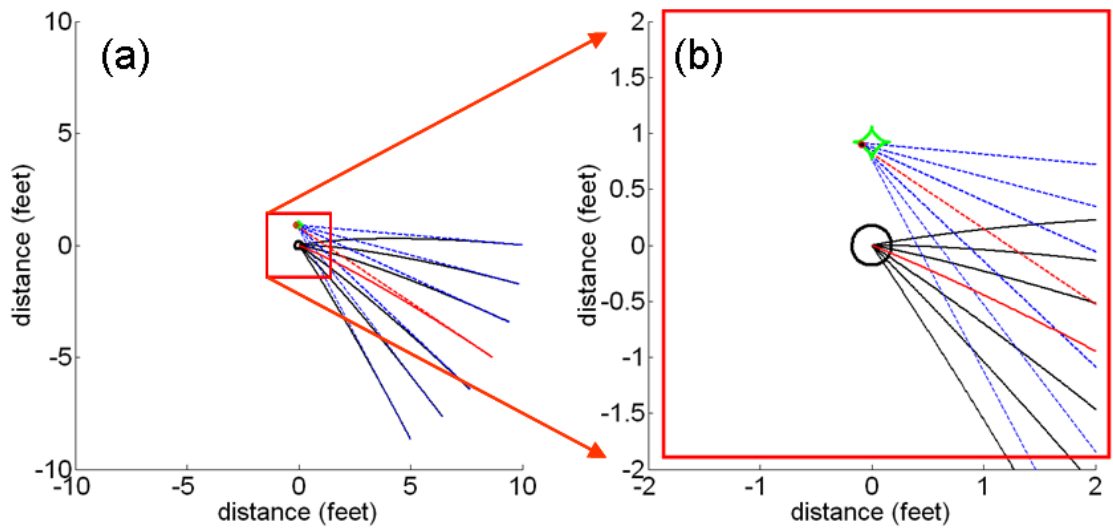

Fig. 22. Putt trajectories for 10 foot putts on a stimp-grade 20 ft-% surface originating in a range of angles centered on the 4 o'clock position and spanning the range from 3 o'clock to 5 o'clock. As can be seen in Fig. 22(b), the target lines all cross very near to one another, sufficiently close that we can call it a target point. The target point is indicated by the red dot. The green diamond shaped structure above the hole is the ensemble of target points for all 10 foot putts on this putting surface.

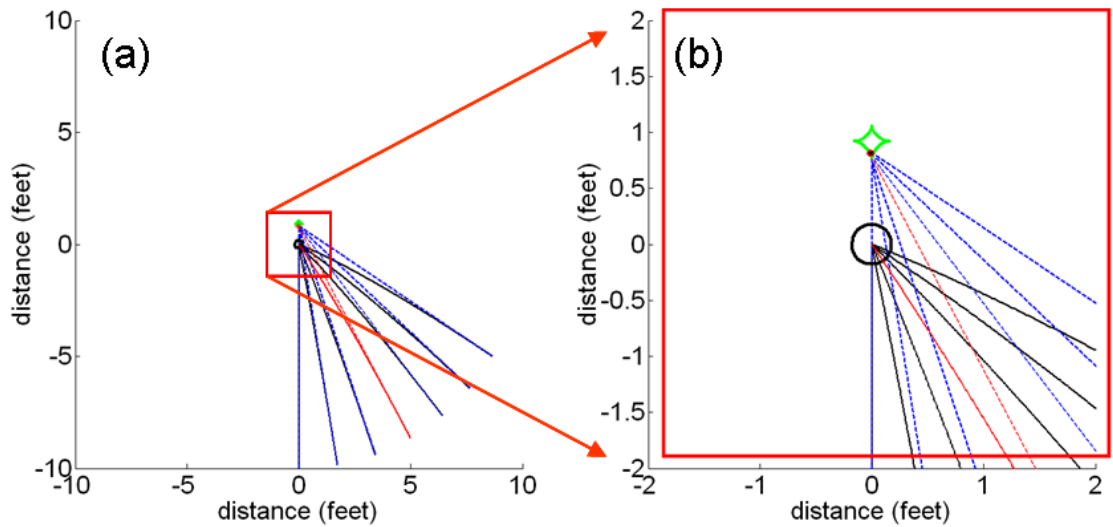

Fig. 23. Putt trajectories for 10 foot putts on a stimp-grade 20 ft-% surface originating in a range of angles centered on the 5 o'clock position and spanning the range from 4 o'clock to 6 o'clock. As can be seen in Fig. 23(b), the target lines all cross very near to one another, sufficiently close that we can call it a target point. The target point is indicated by the red dot. The green diamond shaped structure above the hole is the ensemble of target points for all 10 foot putts on this putting surface.

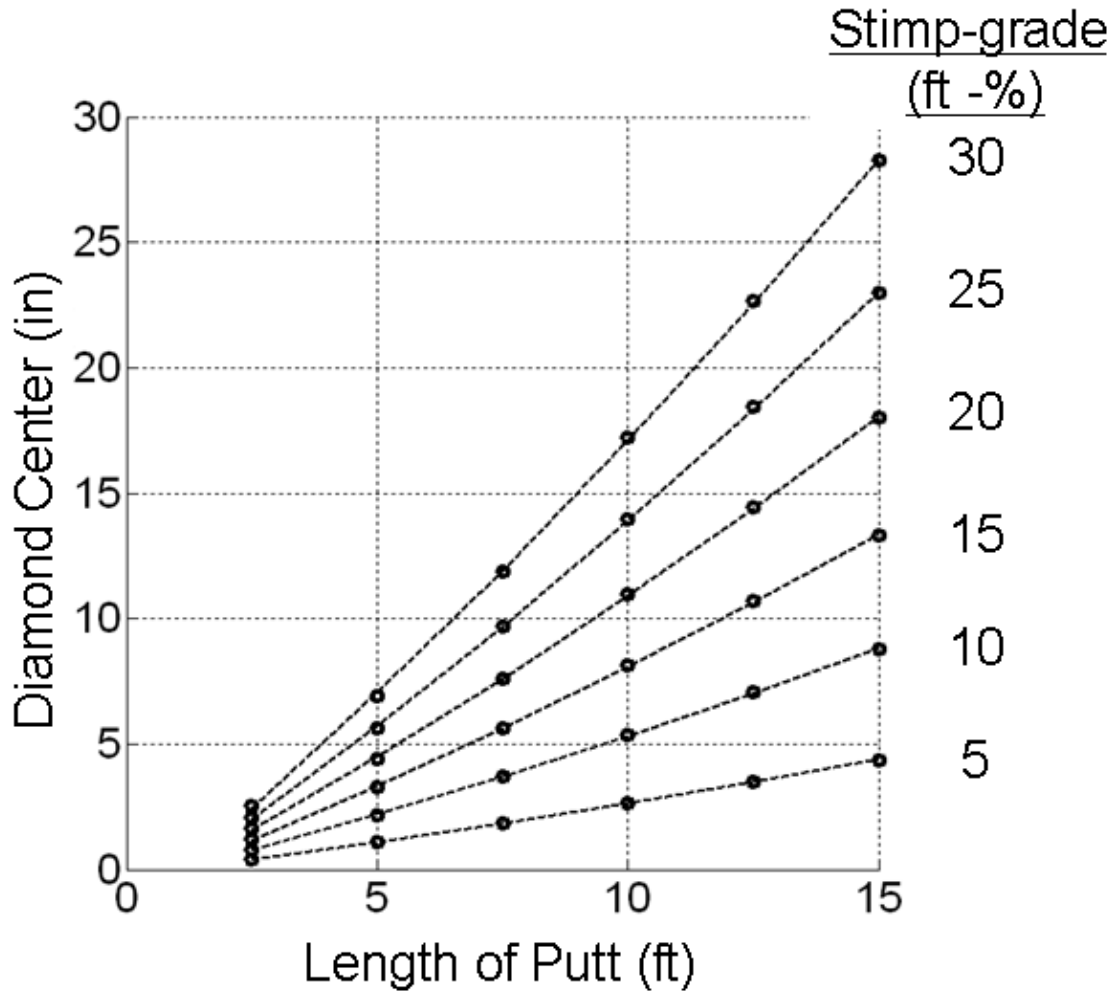

Fig. 24. Center position of the target diamond as a function of putt length for distances from 2.5 thru 15 feet and for values of Stimp-grade product ranging from 5 -30 ft-%. The center position is the distance from the middle of the hole to the middle of the diamond. These curves are calculated for putts with a terminal speed which would have allowed the ball to roll 18 inches beyond the hole. Similar curves could be calculated for any reasonable terminal speed; higher speeds move the curves down and slower speeds move the curves up.

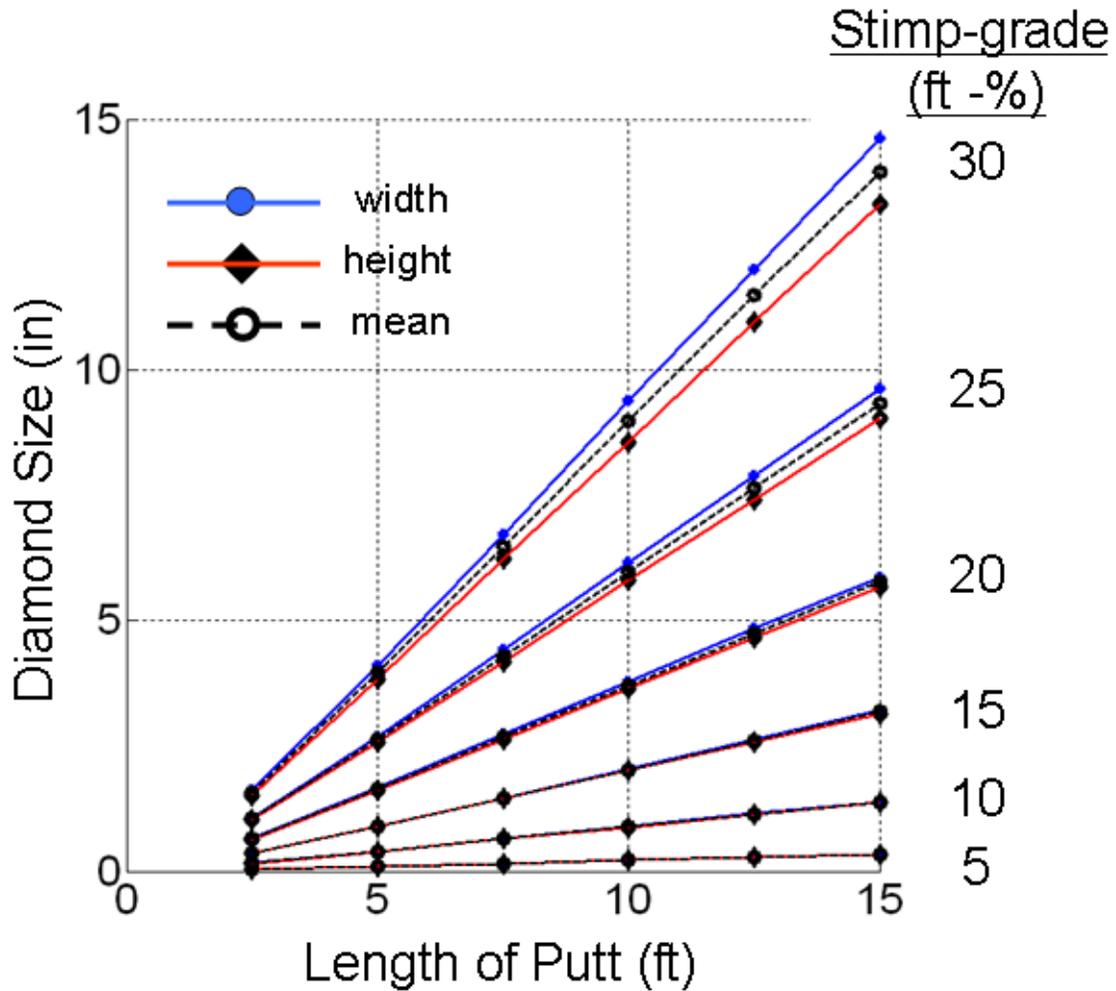

Fig. 25. Width (blue curve) and height (red curve) of the target diamond as a function of putt length for distances from 2.5 thru 15 feet and for values of Stimp-grade product ranging from 5 -30 ft-%. These curves show that the target diamond is very symmetric. The dashed line is the average of the height and width and will be used as a measure of the size scale of the target diamond. These curves are calculated for putts with a terminal speed which would have allowed the ball to roll 18 inches beyond the hole. Similar curves could be calculated for any reasonable terminal speed; higher speeds move the curves down and slower speeds move the curves up.

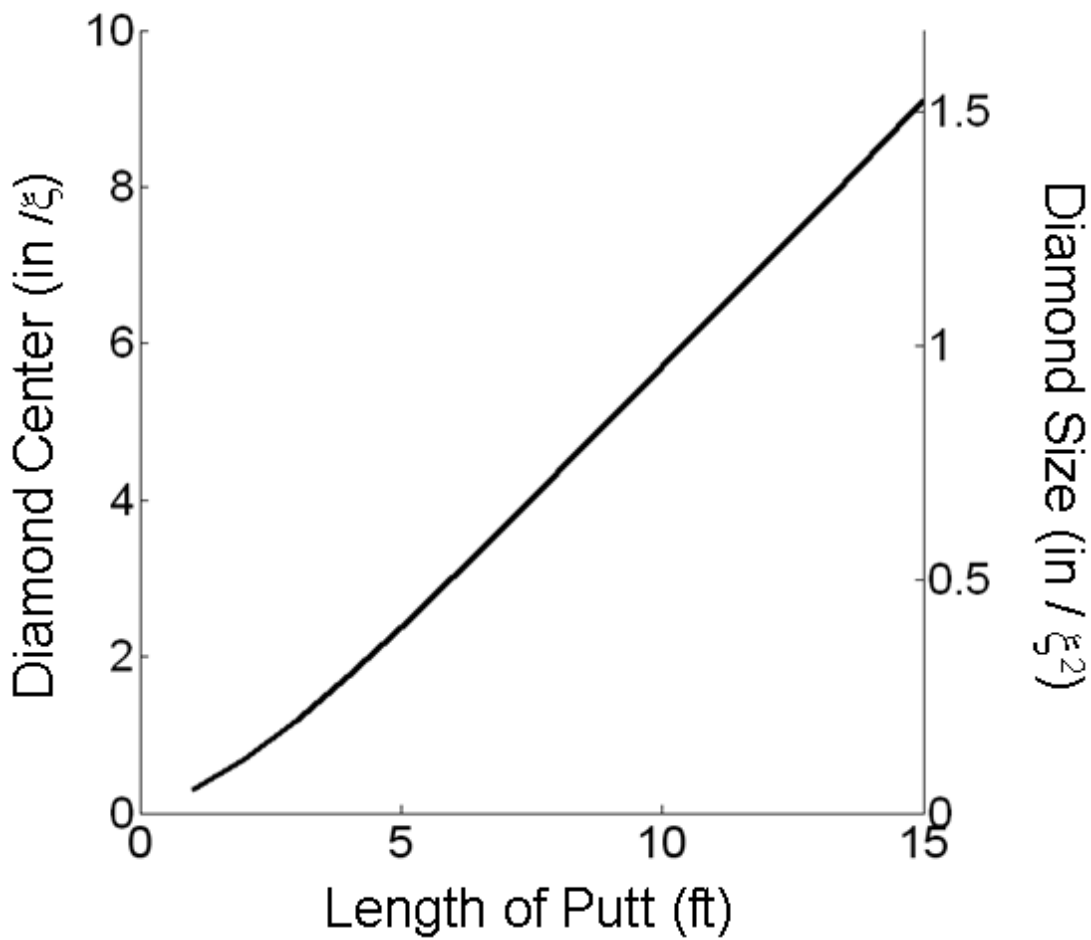

Figure 26: The universal curve from which the dimensions of the target diamond can be determined. The derivation of this curve is described in the text. The left axis indicates the center position of the target diamond normalized by $\xi = d_s\theta/10$. The right axis indicates the size scale (i.e. average of the height and width) of the target diamond normalized by $\xi^2$. Thus, the dependence of the center position on putt length is found by multiplying the left axis by one tenth of the Stimp-grade product. Similarly, the size scale is found by multiplying the right axis by the square of one tenth of the Stimp-grade product. This curve is particular to putts with a terminal speed which would have allowed the ball to roll 18 inches past the hole.

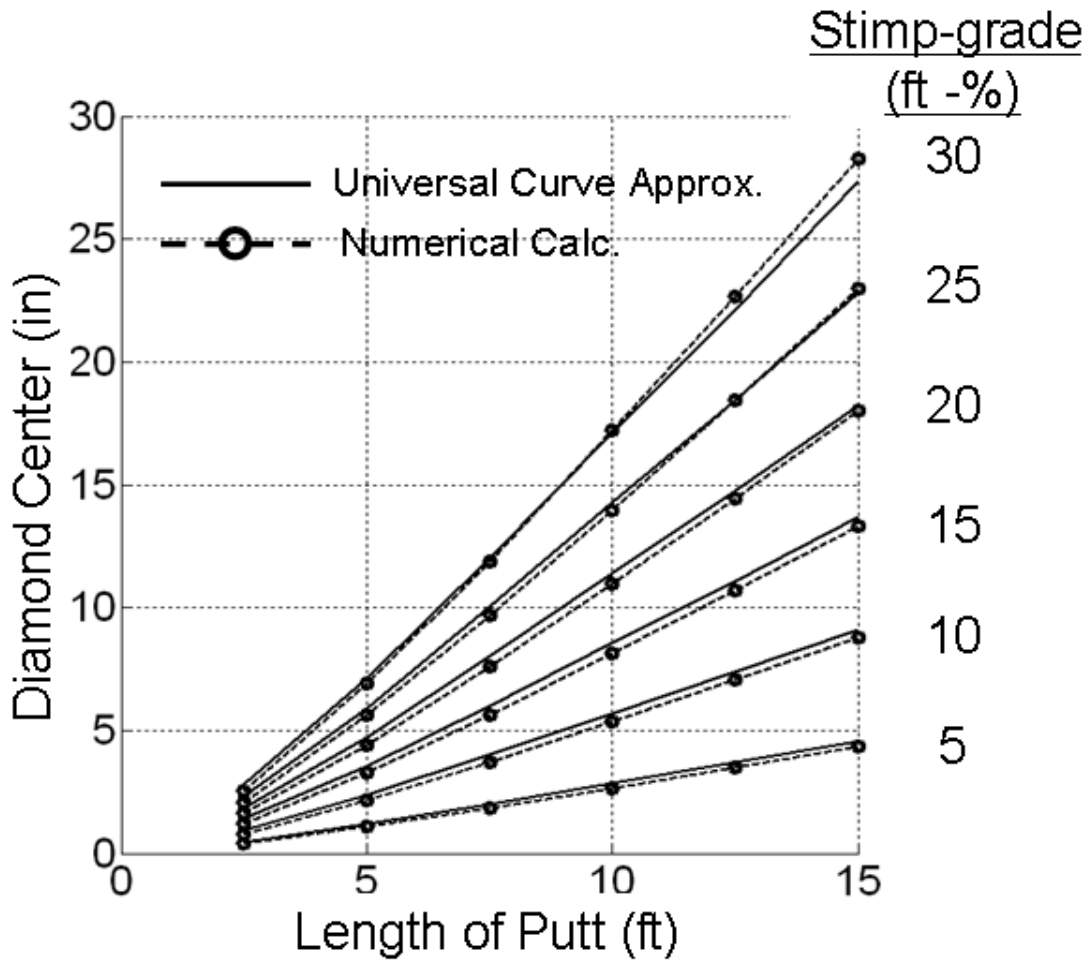

Figure 27. A comparison of the data of Fig. 24 with the universal curve approximation shown in Fig. 26. The universal curve approximation is shown as the black lines and the numerical calculations of Fig. 24 are shown as the open circles connected by dashed lines. The maximum difference between the numerical calculation and the universal curve approximation is less than one inch over the entire data set, and thus the universal curve is a reasonable approximation to the center position of the target diamond.

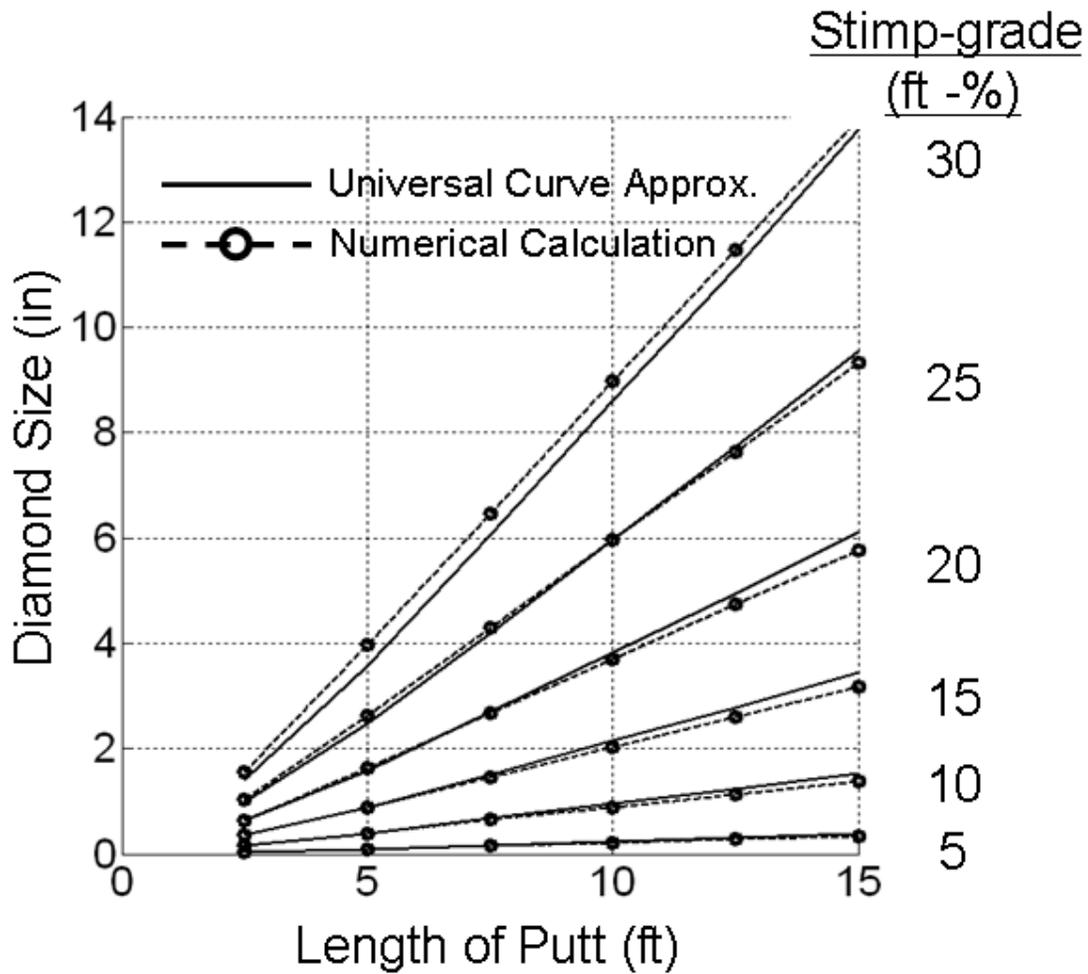

Figure 28. A comparison of the data of Fig. 25 with the universal curve approximation shown in Fig. 26. The universal curve approximation is shown as the black lines and the numerical calculations of Fig. 25 are shown as the open circles connected by dashed lines. The maximum difference between the numerical calculation and the universal curve approximation is less than one inch over the entire data set, and thus the universal curve is also a reasonable approximation to the height and width of the target diamond.